\documentclass[pra,aps,twocolumn,epsfig,showpacs]{revtex4}
\usepackage{amsmath}
\usepackage{amssymb}
\usepackage[dvips]{graphicx}
\usepackage{dcolumn}
\usepackage{bm}
\usepackage[colorlinks=false,dvipdfm]{hyperref} 
\usepackage{setspace}
\usepackage{subfigure}
\usepackage{bm}

\newcommand{\tr}{\mbox{Tr}}
\newcommand{\Exp}{\mbox{Exp}}
\begin{document}

\title{Speed of disentanglement in multiqubit systems under depolarizing channel \footnote{to appear in Annals of Physics }}

\author{Fu-Lin Zhang}\email[Email: ]{flzhang@tju.edu.cn}
\affiliation{Physics Department, School of Science, Tianjin University, Tianjin 300072, China}

\author{Yue Jiang}
\affiliation{Physics Department, School of Science, Tianjin University, Tianjin 300072, China}

\author{Mai-Lin Liang}\email[Email: ]{mailinliang@yahoo.com.cn}
\affiliation{Physics Department, School of Science, Tianjin University, Tianjin 300072, China}

\date{\today}

\begin{abstract}
We investigate the speed of disentanglement in the multiqubit systems under the local depolarizing
channel, in which each qubit is independently coupled to the
environment. We focus on the bipartition entanglement between one qubit
and the remaining qubits constituting the system, which is measured by the negativity.
For the two-qubit system, the speed for
the pure state completely depends on its entanglement.
The upper and lower bounds of the speed for arbitrary two-qubit states, and the necessary conditions for a state achieving them, are obtained.
For the three-qubit system, we study the speed for pure states, whose entanglement properties can be completely described by five local-unitary-transformation invariants.
An analytical expression of the relation between the speed and the invariants is derived.
The speed is enhanced by the the three-tangle which is the entanglement among the three qubits, but reduced by the
the two-qubit correlations outside of the concurrence.
The decay of the negativity can be restrained by the other two negativity with the coequal sense.
The unbalance between two qubits can reduce speed of disentanglement of the remaining qubit in the system, even can retrieve the entanglement partially.
For the $k$-qubit systems in an arbitrary superposition
of Greenberger-Horne-Zeilinger state and W state, the speed depends almost entirely on the amount of the negativity when $k$ increases to five or six.
 An alternative quantitative definition for the robustness of entanglement is presented based on the speed of disentanglement,
  with comparison to the widely studied robustness measured by the critical amount of noise parameter where the
entanglement vanishes.
In the limit of large number of particles, the alternative robustness of the Greenberger-Horne-Zeilinger-type states is inversely proportional to $k$,
and the one of the W states approaches $1/\sqrt{k}$.

\end{abstract}


\maketitle


\section{Introduction}
Entanglement \cite{einstein}, which reveals essentially difference
between quantum and classical mechanics, plays a vital role in many
discussions on the fundamental issues of quantum mechanics. In
recent years, it has been found as a resource of quantum
communication and computation \cite{nielsen}, such as quantum key
distribution \cite{ekert, bennett2}, quantum teleportation
\cite{bennett1} and quantum dense coding \cite{PhysRevLett.69.2881}.

However, it's impossible that a quantum system isolates from its
environment completely, so the system unavoidably loses its
coherence due to interactions with the environment. In composite systems, entanglement
is an manifestation of quantum coherence, which will decay under the influence of
decoherence. This dynamics of entanglement in the open systems has
recently attracted the interest of many researchers \cite{yu1,yonac,qasimi,yu2,roszak2010,laurat,almeida,Salles,simon,aolita,man,liu,zhao}.
It has been shown that in
some cases entanglement can vanish in finite times. This phenomenon of sudden loss of
entanglement is known as \emph{entanglement sudden death} (ESD)
 \cite{zyczkowski2001,yu1,yonac,qasimi,yu2,roszak2010}. The
experimental evidences of the interesting phenomenon have been
reported for atomic ensembles \cite{laurat} and optical setups
\cite{almeida,Salles}.

One of the topics in this direction is about the robustness of entanglement,
 in which the stability of an entangled system is a major concern to the researchers.
Recently, many investigations on the robustness, especially of the multiqubit quantum systems, have been reported.
The concept for the robustness of entanglement has been introduced by Vidal and Tarrach
\cite{vidal1999}  as a
measure of entanglement corresponds to the minimal amount of mixing
with separable states which washed out all entanglement. Then, in some
articles \cite{simon,aolita,man,liu,zhao}, the robustness is defined as the critical amount of
decoherence, where the entanglement entanglement vanishes. In \cite{simon}, the
authors pointed out that, under depolarizing channels, the
robustness of entanglement for multiqubit systems increased with
the qubit number $k$. Aolita \emph{et al.} \cite{aolita} showed that the
time, when the entanglement of a pure Greenberger-Horne-Zeilinger
(GHZ) state system became arbitrarily small and useless as a
resource, was inversely proportional to the particle number $k$,
although its ESD time increased with $k$. In \cite{man}, Man \emph{et al.}
studied the entanglement robustness measured by the
time when an entangled state became completely separable or
remained entangled with a negligible entangled amount, and showed
that entanglement robustness can be enhanced by local operation for
the generalized $k$-qubit GHZ-type states. Zhao and Deng \cite{zhao} studied the multiqubit systems under local depolarizing channel, and found a residual effect on the robustness of a three-qubit system in a superposition
of GHZ state and W state.

Inspired by the results given by Zhao and Deng \cite{zhao},
 the main purpose of this paper is to study the relation between
  the stability of an entangled system and its entanglement properties.
However, the robustness, measured by the critical noise parameters where the ESD occurs, is
determined by the nature of the series of intermediate states from being
entangled to separable under the process of decoherence.
In addition, as shown in \cite{aolita}, the ESD time is irrelevant to assess the robustness of
multiparticle entanglement in some channels.
Therefore, we present the linear response of the entanglement under an infinitesimal noise, namely the \emph{speed of disentanglement} (SoDE), as a quantitative signature of the resistance of an entangled state.
On the other hand, the behavior of the entanglement at the beginning of the decoherence process
 becomes very importance in the application for quantum information, since the coherence time of a
 qubit has been greatly prolonged in some recent schemes \cite{PhysRevLett.105.053201,du2009preserving}.
It is not the first time to concern about the time-derivative of the entanglement,
 for instance, the derivative of the tangle is studied in the investigation about the entanglement flow in multipartite systems \cite{cubitt2005entanglement}.
To exclude the properties of an entangled system affecting the SoDE beside the entanglement,
 we adopt the model of a multiqubit system in which each qubit is coupled to its own depolarizing environment
individually, the same as that in \cite{simon,zhao}. The model is invariable under both the local unitary (LU) transformations and the permutations of the qubits. Furthermore, in the limit of large number of particles, the qualitative behavior of an entangled state is largely independent of the specific
decoherence model \cite{PhysRevA.71.032350}.

As the first trial, in the present paper, we only consider speed of the bipartition entanglement between
 one qubit and the remaining $k-1$ qubits constituting the $k$-qubit system.
We adopt the negativity $\mathcal{N}$ \cite{Vidal,peres,zyczkowski} to measure the entanglement,
which is widely used in recent investigations of the entanglement dynamics in multiqubit system \cite{simon,zhao,PhysRevA.79.022108,PhysRevA.82.032326}.
Compared to other entanglement measurements, such as the concurrence $\mathcal{C}$ \cite{wootters,PhysRevLett.95.040504}, a distinct advantage of the negativity is its computability. Besides, for the arbitrary bipartite pure states, $\mathcal{N}>0$ is the necessary and sufficient condition for entanglement \cite{PhysRevLett.95.040504}. Most of the recent works about the robustness of entanglement in multiparticle system focus on the pure states case.
It is certainly true that our definition of the SoDE is not restricted to the negativity. However, we find the SoDE for concurrence are very close to the one for negativity, in some special cases where the concurrence or its bound can be calculated. The details are given in the conclusion and discussion section.

This paper is organized as follows. In Sec. \ref{model} we make a brief review for the
the decoherence model and the entanglement measurements, and give some general formulas for the SoDE. In Sec. \ref{2qubit}
and \ref{3qubit}, we study the SoDE for
two-qubit arbitrary states and three-qubit pure states, respectively. In Sec.
\ref{mqubit}, we give the result for several special states in multiqubit system, such as GHZ-type state,
W-type state, etc. Conclusion and discussion are made in the last section.

\section{Definitions and general formulas}\label{model}

\subsection{Decoherence Model}
%
%
Under local noise environments, without the interaction between the subsystems in the $k$-party system, the dynamics of each particle
is governed by a master equation which depends on its own environment \cite{PhysRevA.79.022108}.
From the master equation, one can obtain a completely positive trace-preserving map
$\varepsilon_{i}$ which describes the evolution of the corresponding subsystem
\cite{PhysRevA.79.022108}: $\rho_{i}(t)=\varepsilon_{i}\rho_{i}(0),\ (i=1,...,k)$, and
for the whole state $\rho(t)=\otimes_{i=1}^{k} \varepsilon_{i}\rho(0)$. In the
Born-Markovian approximation, the channel can be described by its
Kraus representation \cite{PhysRevA.79.022108,zhao,kraus} as
\begin{eqnarray}
\varepsilon_{i}\rho_{i}(0)=\sum_{j=0}^{M}E_{ji}\rho_{i}(0)E_{ji}^{\dag}
\end{eqnarray}
where $E_{ji},j=0,...,M$ are the so-called Kraus operators needed to
completely characterize the channel.

In detail, for the one-qubit quantum system, the Kraus operators of the partially depolarizing channel can be expressed as
\begin{eqnarray}
E_0=\sqrt{1-p'}I,\ \  E_j=\sqrt{\frac{p'}{3}}\sigma_j,
\end{eqnarray}
where $I$ is the unit matrix,  $p'=\frac{3p}{4}$ with $p$ being the depolarization parameter, and $\sigma_{j}\ (j=1,2,3)$ are the corresponding
Pauli matrices. In the
Bloch sphere representation for the qubit density operator, the map can be written as
$\varepsilon \rho =\frac{1}{2}(I+
s\textbf{r}\cdot\bm{\sigma})$, where $s=1-p$.
Following the Ref. \cite{dur}, we
consider the depolarization parameter $s=s(t)= e^{-\kappa t}$, where $\kappa$ is a decay constant determined by the
strength of the coupling to the environment and $t$ is the
interaction time. Without loss of generality, we set $\kappa=1$ and
$s= e^{-t}$ in the present work.

\subsection{Entanglement measurements and invariants}
Several concepts have been presented to quantify or describe the entanglement in quantum systems, such as the entanglement of formation
\cite{wootters}, the entanglement cost \cite{Bennett3}, the
distillable entanglement \cite{Bennett4}, the relative entropy of
entanglement \cite{Vedral}, and so on.
Here, we briefly review the concurrence \cite{wootters} and negativity \cite{Vidal} for bipartition entanglement, and the entanglement invariants for three-qubit pure states, which are utilized in this paper.

The \emph{negativity} as a measure of entanglement was introduced by
\.{Z}yczkowski et al. \cite{zyczkowski}, base on
positive partial transpose (PPT) criterion\cite{peres}, which is necessary for separability, but sufficient only for
$2\otimes2$ and $2\otimes3$ systems \cite{horodecki1996}. For a bipartite system described by the density matrix  $\rho$,
the negativity is defined as  \cite{zyczkowski,peres}
\begin{eqnarray}\label{Neg}
\mathcal{N}(\rho)= 2   \sum_{j} |\lambda_j|,
\end{eqnarray}
where $\lambda_j$ are the negative eigenvalues of $\rho^{T}$ and
$T$ denotes the partial transpose operation on one of the subsystems.
For a pure $m\otimes n\ (m\leq n)$ state in the standard Schmidt form
\begin{eqnarray}\label{mnstate}
| \psi \rangle= \sum_{i} a_i   |\mu_i \nu_i \rangle
\end{eqnarray}
where $a_i\in[0,1] \ (i=1,...,m) $ are the Schmidt coefficients satisfying $\sum_{i} a_i^2 =1$, $|\mu_i \rangle$
and $|\nu_i \rangle$ are the orthonormal basis of the two subsystems,
the negativity is given by \cite{PhysRevLett.95.040504}
\begin{eqnarray}\label{NegPure}
\mathcal{N}( | \psi \rangle )= \biggr( \sum_{i} a_i \biggr)^2-1,
\end{eqnarray}
from which one can find the fact that, $\mathcal{N}>0$ is the necessary and sufficient condition for the entanglement in pure states.

The \emph{concurrence} originated in the investigation of the entanglement of formation \cite{wootters}.
The concurrence of a pure state $| \psi \rangle$ in a bipartite system is given
by \cite{PhysRevLett.95.040504}
\begin{eqnarray} \label{ConP}
\mathcal{C}(| \psi \rangle) =\sqrt{2(1- \tr  \rho_A^2)}= \sqrt{2(1- \tr \rho_B^2)},
\end{eqnarray}
where $\rho_A= \tr_B | \psi \rangle \langle \psi |$ is the partial
trace of $| \psi \rangle \langle \psi |$ over subsystem $B$, and
$\rho_B$ has a similar meaning. For a mixed state, the concurrence
is defined as the average concurrence of the pure states of the
decomposition, minimized over all decompositions of $\rho =
\sum_{j}p_{j} | \psi_j \rangle  \langle \psi_j |$,
\begin{eqnarray}\label{ConM}
\mathcal{C}(\rho)= \min \sum_{j}p_{j}\mathcal{C}(| \psi_j \rangle).
\end{eqnarray}
For the two-qubit case, it is equivalent to the entanglement of formation \cite{wootters} and can be expressed explicitly as
\begin{eqnarray}\label{ConEx}
\mathcal{C}(\rho)= \max
\{0,\lambda_{1}-\lambda_{2}-\lambda_{3}-\lambda_{4}\},
\end{eqnarray}
in which $\lambda_{1},...,\lambda_{4}$ are the square roots of the eigenvalues of the
operator $R=\rho (\sigma_{y} \otimes \sigma_{y} ) \rho^{*}
(\sigma_{y} \otimes \sigma_{y} )$ in decreasing order and
$\sigma_{y}$ is the second Pauli matrix.

For an $m\otimes n\ (m\leq n)$ arbitrary quantum state $\rho$, the negativity
is proved to provide a lower bound of the concurrence as \cite{PhysRevLett.95.040504}
\begin{eqnarray}\label{NboundC}
\mathcal{C}(\rho)\geq \sqrt{\frac{2}{m(m-1)}}\mathcal{N}(\rho).
\end{eqnarray}
In the present work, we focus on the bipartite entanglement between the $j$th qubit and the remain $k-1$ qubits constituting the $k$-qubit system.
Under the two measurements, it is denoted by $\mathcal{N}_j$ and $\mathcal{C}_j$ respectively.
This corresponds to the case of $m=2$ in (\ref{NboundC}), where the relation  reduces to $\mathcal{C}_j(\rho)\geq \mathcal{N}_j(\rho)$. And for the $k$-qubit pure states $\mathcal{C}_j(| \psi \rangle)=\mathcal{N}_j(| \psi \rangle)$, which can be obtained directly from their definitions in (\ref{NegPure}) and (\ref{ConP}).


The \emph{entanglement invariants} \cite{sudbery,acin1,acin2} of three-qubit pure states
 are five linear independent polynomials, which is invariance under LU transformations.
For the three-qubit pure state $\rho=| \psi \rangle \langle \psi |$, they are given by
\begin{eqnarray}\label{I3qubit}
\mathcal{I}_i&=&\tr \rho_i^2 =1-\frac{\mathcal{N}_{i}^2}{2}, \nonumber\\
\mathcal{I}_4&=&3\tr[(\rho_i\otimes\rho_j)\rho_{ij}]-\tr \rho_i^3 -\tr \rho_j^3 , \\
\mathcal{I}_5&=&\tau^2=(\mathcal{N}_{i}^2-\mathcal{C}_{ij}^2-\mathcal{C}_{ik}^2)^2 , \nonumber
\end{eqnarray}
where $\rho_i=\tr_{jk} \rho \ (i,j,k=1,2,3)$ are the one-particle density matrices, $\rho_{ij}=\tr_{k} \rho$ are the
two-particle density matrices, and $\mathcal{N}_i$ and $\mathcal{C}_{ij}$ are the negativity and concurrence of the
reduce states with the corresponding subscripts.
The invariants $\mathcal{I}_{1,2,3}$ are equivalent to the one-qubit linear entropies, which
characterize the entanglement between one qubit and the the remaining two qubits.
The last invariant $\mathcal{I}_5$ is equivalent to the three-tangle
$\tau$, which describes the whole entanglement of the
three-qubit system \cite{coffman,lohmayer}. The remaining invariant $\mathcal{I}_4$
is related to the relative entropy of the two-qubit
state $\rho_{ij}$ relative to the product state $\rho_i\otimes\rho_j$, and is a second measure
of the entanglement of in the reduced states of $\rho_{ij}$,
 independent of the concurrence $\mathcal{C}_{ij}$ which can be determined by $\mathcal{N}_{1,2,3}$ and $\tau$ \cite{sudbery}.
%
%

\subsection{SoDE and perturbation approach}

Under the local noise channels, the SoDE we focus on in the present article is the
one between the $i$th qubit and the other part of the whole system. Utilizing the
entanglement measure negativity, it can be expressed as
\begin{eqnarray}\label{eta}
\eta_i (\rho)=\biggr|\frac{d \mathcal{N}_i(\rho_s)}{d t}\biggr|_{t=0},
\end{eqnarray}
where
$\rho_s=\otimes^{k}_{j=1}\varepsilon_j \rho $ is the final state with $\rho$ being the initial one.
To compare with the widely utilized definition of the robustness by the ESD noise parameter,
 we present an alternative definition
of robustness of entanglement, which is
\begin{eqnarray}\label{aRob}
\mathcal{R}_{\eta_i}=1-\Exp(-T_i^*),
\end{eqnarray}
where $T_i^*=\frac{\mathcal{N}_i}{\eta_i}$ is a characteristic time of the
disentanglement.

We find that the SoDE $\eta_{i}$ can be derived with the
perturbation theory \cite{pertubation}. We called the method as the \emph{perturbation approach}.
In quantum mechanics, the
perturbation theory is applied to the systems whose Hamiltonian can be divided into
\begin{eqnarray}
H=H_0 + \epsilon W,
\end{eqnarray}
where $\epsilon$ is the smallness parameter, and $H_0$ is the easily solvable unperturbed Hamiltonian.
The eigenvalues and eigenstates of $H_0$ are given by
\begin{eqnarray}
H_0 |\phi_k \rangle = E_k^{(0)} |\phi_k \rangle.
\end{eqnarray}
The eigenvalues of the Hamiltonian $H$ can be expanded in powers of the perturbation parameter $\epsilon$,
\begin{eqnarray}
E_k = E_k^{(0)} + \epsilon E_k^{(1)} + \epsilon^2 E_k^{(2)}+ ...
\end{eqnarray}
For the state $ |\phi_k \rangle$ without degeneracy, the first approximation is given by
\begin{eqnarray}\label{nonG1st}
  E_k^{(1)} = \langle \phi_k| W |\phi_k\rangle.
\end{eqnarray}
When the energy level $E_k^{(0)}$ is $d_k$-fold degenerate, with the eigenstates $|\phi_k^j \rangle$, ($j=1,2...d_k$),
 the corresponding eigenvalues of $H$ in the first approximation are
\begin{eqnarray}\label{Gener}
E_{k, \alpha} = E_k^{(0)} + \epsilon \beta_{k,\alpha},\ \ \ \alpha=1,2...d_k,
\end{eqnarray}
where $\beta_{k,\alpha}$ is the $\alpha$-th eigenvalues of the $d_k \times d_k$ matrix $W_k$ with the elements $W_{k,ij}= \langle \phi_k^i |W|\phi_k^j\rangle$.
One can notice that, when $d_k=1$, $\beta_{k,1}$ becomes the result for the non-degenerate case in Eq. (\ref{nonG1st}).
And, the trace of $W_k$ satisfies
\begin{eqnarray}\label{trace}
\tr W_k=\sum_{\alpha=1}^{d_k} \beta_{k,\alpha}= \sum_{j=1}^{d_k} \langle \phi_k^j |W|\phi_k^j\rangle.
\end{eqnarray}

Through the noise channel in an infinitesimal time
 $d t$, the state of a system can be written as
\begin{eqnarray}
\rho_s=\rho+\sigma dt,
\end{eqnarray}
where  $\rho$ is the initial state, $\sigma$ is an Hermitian
operator with $\tr \sigma=0$. The negativity of
the final state is given by $\mathcal{N}_i(\rho_s)=\mathcal{N}_i(\rho)-\eta_i dt$.
Considering the partial transposed state $\rho_{s}^{T_{i}}=\rho^{T_{i}}+\sigma^{T_{i}} dt$
 as the Hamiltonian of a quantum system, and $dt$ as the smallness parameter,
  we find the SoDE $\eta_{i}$ is determined by the first-order response of the eigenvalues
of $\rho^{T_{i}}$ under the perturbation $\sigma^{T_{i}} dt$.
 Since the negativity is defined as the sum of the negative eigenvalues of the partial transposed density matrix,
only the eigenvectors of $\rho^{T_i}$ with the zero and the negative eigenvalues contribute to
the speed $\eta_i$. Namely, the negativity of $\rho_s$ is
\begin{eqnarray}
\mathcal{N}_i(\rho_s)= -2 \sum_j \lambda_{j,-} - 2  \sum_j \lambda_{j,-}^{(1)} dt - 2  \sum_l \lambda_{l,0,-}^{(1)} dt,
\end{eqnarray}
where $ \lambda_{j,-}$ are the negative eigenvalues of $\rho^{T_i}$ and $\lambda_{j,-}^{(1)}$ are the first-order response under the perturbation $\sigma^{T_{i}} dt$, and $\lambda_{l,0,-}^{(1)}$ are the negative responses of the zero eigenvalue. Therefore, the speed can
be divided into two parts as
\begin{eqnarray}\label{pert}
\eta_i=\eta_i^{(-)}-\eta_i^{(0)},
\end{eqnarray}
where $\eta_i^{(-)}=2  \sum_j \lambda_{j,-}^{(1)} $ and $\eta_i^{(0)}=-2 \sum_l \lambda_{l,0,-}^{(1)}$.
According to the relations in Eqs. (\ref{nonG1st}), (\ref{Gener}) and (\ref{trace})
the first term can be expressed as
 \begin{eqnarray}\label{etam}
\eta^{(-)}_i=2\sum_k{\langle
\psi_k^-|\sigma^{T_i}|\psi_{k}^-\rangle},
\end{eqnarray}
where $|\psi_{k}^-\rangle$, $(k=1,2...d_-)$ are the eigenvectors
 of $\rho^{T_{i}}$ with negative eigenvalues.
And the second one corresponds to the eigenvectors
$|\psi_{j}^0\rangle$ $(j=1,2...d_0)$ with the zero eigenvalues is
$\eta_i^{(0)}= \sum_l | \lambda_{l,0}^{(1)} | -   \sum_l \lambda_{l,0}^{(1)} $, where $\lambda_{l,0}^{(1)}$ are the first-order
responses of the zero eigenvalues.
It is equivalent to
\begin{eqnarray}\label{eta0}
\eta^{(0)}_i=\| \sigma^{T_i}_0 \| -\tr \sigma^{T_i}_0,
\end{eqnarray}
where $\sigma^{T_i}_0 $ is the $d_0 \times d_0$ matrix with the
elements  $\sigma^{T_i}_{0,mn} =\langle
\psi_m^0|\sigma^{T_i}|\psi_{n}^0\rangle $, and $\| \cdot \|$ stands for the trace norm defined by $\| G \|= \tr \sqrt{G G^{\dag}}$.

\section{Two-qubit arbitrary states} \label{2qubit}

First we explore the connection between the SoDE of a two-qubit quantum system and its entanglement properties.
Since the entanglement is invariant under permutations of the two qubits,
 we omit the subscripts of the entanglement and the SoDE.

\subsection{Pure states}
 The pure states of a two-qubit system are always equivalent to
\begin{eqnarray}\label{PureTh}
| \psi (\theta) \rangle = \cos \theta | 00 \rangle + \sin \theta |
11 \rangle,\ \ \ \theta \in [0,\pi/4],
\end{eqnarray}
under LU transformations, with the entanglement
$\mathcal{N}[| \psi (\theta) \rangle]=\mathcal{C}[| \psi (\theta) \rangle]= \sin 2\theta$.
The evolution of the negativity for the states under the local depolarizing channel can be
calculated directly. But we are willing to give the the following steps to demonstrate the
perturbation approach introduced in the above section.
The partial transposed density matrix of the pure state (\ref{PureTh}) is given by
\begin{eqnarray}\label{purep}
\rho^{T}=\left[\begin{array}{cccc}
 \cos^2 \theta  & 0 & 0 & 0 \\
 0 & 0 & \frac{   \sin 2 \theta}{2}   & 0 \\
 0 &  \frac{ \sin 2 \theta}{2}   & 0 & 0 \\
 0 & 0 & 0 &  \sin^2\theta
\end{array}
\right],
\end{eqnarray}
which has no zero eigenvalue, and a eigenvector with the negative eigenvalue being $| \phi^{-} \rangle = (|01\rangle-|10\rangle)/\sqrt{2}$.
Under the depolarizing
channel in a  finite time $t$, the partial transposed
density matrix becomes
\begin{eqnarray}\label{pureps}
\rho_{s}^{T}=\left[
\begin{array}{cccc}
 \frac{1+s^2+2 s \cos 2 \theta}{4} & 0 & 0 & 0 \\
 0 & \frac{1-s^2}{4}  & \frac{s^2 \sin 2 \theta}{2}  & 0 \\
 0 & \frac{s^2 \sin 2 \theta}{2}  & \frac{1-s^2}{4}  & 0 \\
 0 & 0 & 0 & \frac{1+s^2-2 s \cos 2 \theta}{4}
\end{array}
\right], \ \
\end{eqnarray}
where $s=e^{-t}$.
The perturbation matrix can be derived as
\begin{eqnarray}
\sigma^{T}=\frac{d  \rho_{s}^{T}}{d t}\biggr|_{t=0}=\left[
\begin{array}{cccc}
 - \cos^2  \theta & 0 & 0&0 \\
 0 & 1/2 & - \sin 2 \theta &0\\
 0 & - \sin 2 \theta & 1/2&0\\
 0&0&0& - \sin^2  \theta
\end{array}
\right].
\end{eqnarray}
Substituting $| \phi^{-} \rangle $ and $\sigma^{T}$ into (\ref{pert}) and (\ref{etam}), we obtain the SoDE for the two-qubit pure states as
\begin{eqnarray}\label{SoDE2bit}
\eta=2 \mathcal{N}+1,
\end{eqnarray}
which is a linear function of the entanglement.
 It is interesting to notice $\eta \rightarrow 1$ when $ \mathcal{N} \rightarrow 0$,
  which indicates the ESD occurs in the pure state of this model even with a slight entanglement.
In such a situation, the ESD time equals to the characteristic time $T_{ESD} \rightarrow T^* \rightarrow \mathcal{N} $,
 and the robustness in (\ref{aRob}) is given by $\mathcal{R}_{\eta} \rightarrow \mathcal{N} $ approaching the result given in \cite{zhao}.

\subsection{Frontier states}

For the arbitrary two-qubit states, since their entanglement can't be completely described by one entanglement measure,
 it is very difficult to derive an analytical expression for the relation between the SoDE and the entanglement properties.
Therefore we explore the bounds of SoDE for a given value of negativity and the states achieve the bounds, which is called
 as the \emph{frontier states} in this article.

\begin{figure}
\subfigure[]{\includegraphics[width=7cm]{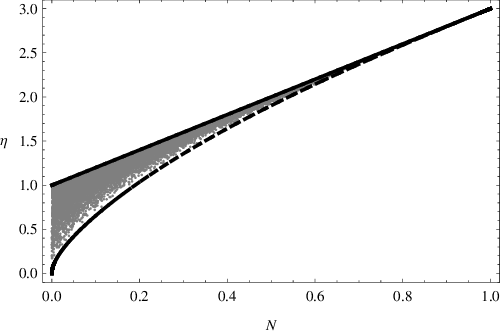}}
\subfigure[]{\includegraphics[width=7cm]{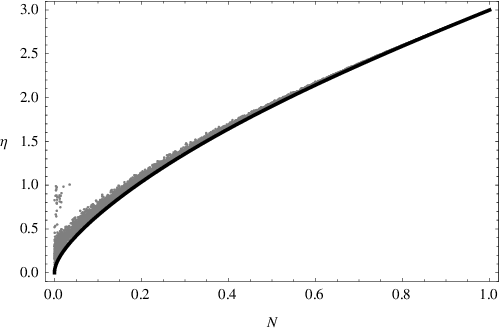}}
 \caption{Plots of the relations between the SoDE and negativity of randomly generated two-qubit states: (a)
 30000 randomly generated states and the curves showing the upper bound (solid) and the lower bound (dashed) of $\eta$;
(b) 5000 randomly generated states weighted with the curve of the lower bound.
 }
\label{fig1}
\end{figure}

Following the approach in \cite{Munro,wei}, we generate randomly a great number of
two-qubit states and plot them in the $\eta$-$\mathcal{N}$ plane as shown in Fig. \ref{fig1} (a).
We fortunately find the pure states are the frontier states with the maximum of SoDE.
To derive the frontier states achieving the lower bound, we begin with the ansatz states
\cite{Munro,wei}
%
%
\begin{eqnarray}\label{ansatz}
\rho_{ansatz}=\left[
\begin{array}{cccc}
 x+\frac{\gamma}{2} & 0 & 0&\frac{\gamma}{2} \\
 0 & a & 0 &0\\
 0 & 0 & b&0\\
 \frac{\gamma}{2}&0&0&y+\frac{\gamma}{2}
\end{array}
\right],
\end{eqnarray}
where $x,y,a,b,\gamma\geq0$ and $x+y+a+b+\gamma=1$.
 The region of the ansatz states in the $\eta$-$\mathcal{N}$ plane is the same as the one of the arbitrary states.
By trying to adjust the constrains imposed on the ansatz states without leaving the lower bound,
 we find the frontier states $\rho_{m}$ on the lower bound when $x=y=b=0$, namely
 \begin{eqnarray}\label{rhom}
\rho_{m}= \gamma | \psi(\pi/4)\rangle \langle \psi(\pi/4) | +(1-\gamma) | 01\rangle \langle01 |,
\end{eqnarray}
with the pure states $| \psi(\pi/4)\rangle$ defined in (\ref{PureTh}).
To verify the frontier state $\rho_{m}$, we randomly generate weighted random states, namely the mixtures of random states
and $\rho_{m}$ with random weights, and plot them in the $\eta$-$\mathcal{N}$ plane. As shown in Fig. \ref{fig1} (b), the region of physically acceptable
states is encircled perfectly by the curve of
the state $\rho_{m}$.
%
The results of the two families of frontier states presented the lower and
upper bounds of the SoDE in two-qubit system as
\begin{eqnarray}\label{BoundsSoDE}
\frac{\mathcal{N}^2+\sqrt{2\mathcal{N}(\mathcal{N}+1)}}{1+2 \mathcal{N}-\sqrt{2\mathcal{N}(\mathcal{N}+1)}} \leq \eta \leq 2\mathcal{N}+1.
\end{eqnarray}
When $\mathcal{N} \rightarrow 0$, the minimum of the SoDE approaches $\sqrt{2\mathcal{N}}$, and the ESD time and the characteristic time $T_{ESD} \rightarrow T^* \rightarrow \sqrt{\mathcal{N}/2}$.

According with the results in \cite{verstraete}, the two families of frontier states exactly is the ones in the comparison of the concurrence and negativity.
Specifically, the pure states which has the maximal SoDE achieves the minimal concurrence when the negativity is fixed,
 and the states $\rho_{m}$ with the minimum of SoDE has the maximum of concurrence for a given negativity.
This results indicate that the concurrence reduces the attenuation of the negativity.

\begin{figure}
\subfigure[]{\includegraphics[width=7cm]{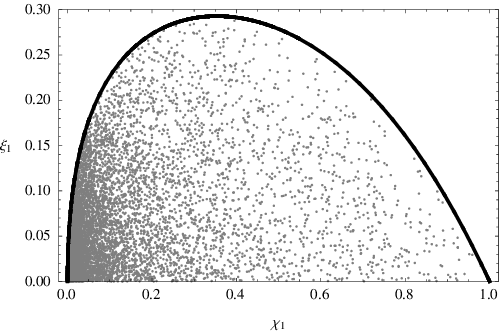}}
\subfigure[]{\includegraphics[width=7cm]{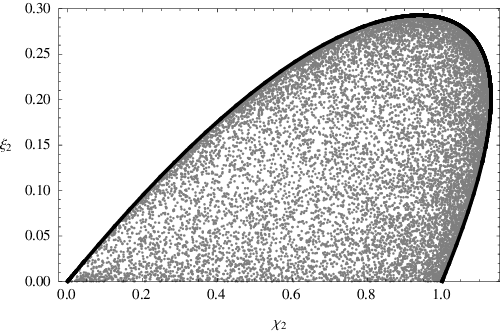}}
 \caption{Plots of 30000 randomly generated states in (a) the
$\xi_{1}-\chi_{1}$ plane and (b) the $\xi_{2}-\chi_{2}$ plane.
The solid curves are the results of $\rho_m$ in (a) and $\rho_k$ in (b).
 }
\label{fig2}
\end{figure}

Then, a question arises: Is there a definite link between the extreme cases in the two topics?
To explore the relations, we introduce four non-negative quantities as
$\xi_{1}=\mathcal{C}-\mathcal{N}$, $\chi_{1}=2\mathcal{N}+1-\eta$, $\xi_{2}=\mathcal{N}-\mathcal{N}^{min}$ and $\chi_{2}=\eta-\eta^{min}$,
where $\eta^{min}$ is the lower bound of the SoDE in (\ref{BoundsSoDE}) and $\mathcal{N}^{min}=\sqrt{\mathcal{C}^2+(1-\mathcal{C})^2}+\mathcal{C}-1$
is the minimal negativity for a fixed concurrence satisfied by the frontier states $\rho_{m}$.
Plotting the randomly generated states in the $\xi_{1}$-$\chi_{1}$ and $\xi_{2}$-$\chi_{2}$ planes as shown in Fig. \ref{fig2}, in the same approach for $\eta$-$\mathcal{N}$, we obtain the frontier states in the two planes.
They are the states $\rho_m$ for the first case, and
\begin{eqnarray}\label{rhok}
\rho_{k}= \gamma | \psi(\pi/4)\rangle \langle \psi(\pi/4) | + a( | 01\rangle \langle01 |+| 10 \rangle \langle10 |),
\end{eqnarray}
with $\gamma+2a =1$, for the second, which are shown by the solid lines in Fig. \ref{fig2}.
By straightforward calculations, one can find that, $\xi_1(\rho_m)$ is a single-valued function of $\chi_1(\rho_m)$, and equally true for $\xi_2(\rho_k)$ and $\chi_2(\rho_k)$. Hence, the states $\rho_m$  has the maximums of $\xi_1$ for a given $\chi_1$.
When $\chi_1(\rho_m)=0$, it can be derived that $\xi_1(\rho_m)=0$. Therefore, for arbitrary states, $\chi_1=0 \Rightarrow \xi_1=0$.
In other words, the concurrence reaching the minimum is a necessary condition for SoDE with the maximum.
Similarly, one can find that the concurrence reaching the maximum is a necessary condition for SoDE with the minimum.
These demonstrates that there  are  other factors that affect the SoDE in two-qubit system in addition to the concurrence and negativity.

\subsection{Two-parameter states}
To study the influence on the SoDE by the concurrence and other quantities related with entanglement,
we consider three classes of two-parameter mixed states in the similar form as the ansatz states (\ref{ansatz}).

\begin{figure}
\subfigure[]{\includegraphics[width=5cm]{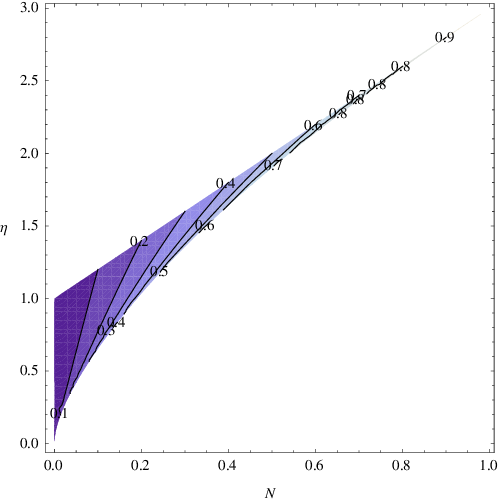}}
\subfigure[]{\includegraphics[width=5cm]{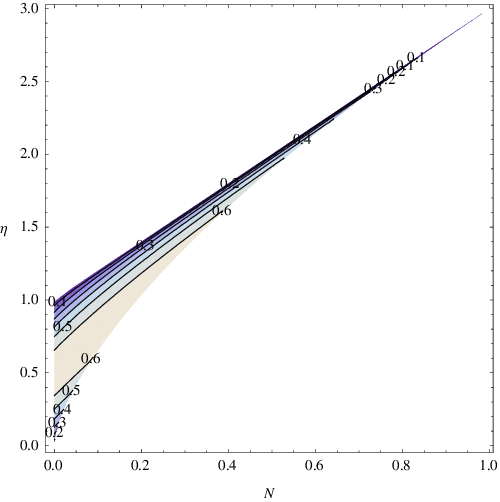}}
\subfigure[]{\includegraphics[width=5cm]{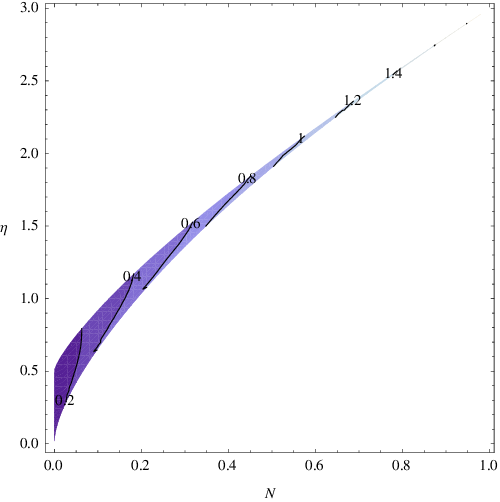}}
 \caption{ Plots of the relations between the SoDE and the negativity of the states (a) $\rho_C$, (b) $\rho_{S_L}$ and (c) $\rho_{I_{tot}}$, with the
 contours of the concurrence $\mathcal{C}$, the linear entropy $\mathcal{S}_L$ and the quantum mutual information $\mathcal{I}_{tot}$ respectively.
 }
\label{fig3}
\end{figure}

%
As an analytical example to show the relation between the SoDE and the concurrence,
the first two-parameter class of states we studied is
\begin{eqnarray}\label{rhoC}
\rho_{C}= \gamma | \psi(\pi/4)\rangle \langle \psi(\pi/4) | + a| 01\rangle \langle01 |+ b  | 10 \rangle \langle10 | ,
\end{eqnarray}
with $\gamma, a, b \in[0,1]$ and $\gamma + a+b =1$, which is the ansatz state with $x=y=0$.
When $a=b$, it returns to the states $\rho_k$ in (\ref{rhok}), whose concurrence and negativity have the same values.
When $ab=0$, it reduces the states with the maximal concurrence, such as $\rho_m$ in (\ref{rhom}).
Its SoDE can be expressed in terms of $\mathcal{N}$ and $\mathcal{C}$ as
\begin{eqnarray}
\eta=2\mathcal{N}+1-\frac{2(1-\mathcal{C})(\mathcal{C}-\mathcal{N})(1+\mathcal{N})}{\mathcal{N}^2-\mathcal{C}^2+2\mathcal{C}(1+\mathcal{N})},
\end{eqnarray}
which can returns to the bounds in (\ref{BoundsSoDE}) in the two cases of $a=b$ and $ab=0$ respectively.
In the Fig. \ref{fig3} (a), one can notice in this family of states, when the other entanglement measure is fixed, the SoDE increases as the negativity increases, but decrease with the concurrence increasing.
%
%
%

Considering the frontier states $\rho_m$ in (\ref{rhom}) are precisely
the \emph{maximally entangled mixed states} in the region of $\mathcal{C} \geq 2/3$ \cite{Munro},
we attempt to find the relation between the SoDE and the the degree of mixture,
 measured by the linear entropy as $\mathcal{S}_L(\rho)=\frac{4}{3}(1-\tr \rho^2 )$.
Since, for a given concurrence the pure states has the minimal linear entropy,
we choose the hybrid states of (\ref{PureTh}) and (\ref{rhom}) as
\begin{eqnarray}\label{rhoSL}
\rho_{S_{L}}=\gamma|\psi(\theta)\rangle\langle\psi(\theta)|+(1-\gamma)|01\rangle\langle01|.
\end{eqnarray}
In Fig. \ref{fig3} (b), one can notice for a fixed mixture in these states, the relation between the SoDE and negativity is approximately linearity.
However, the SoDE does not clearly relate with the linear entropy  $\mathcal{S}_L$.
%

Rather than the mixture of the whole state, we are interested in the quantities describing the
relations between the subsystems. The total correlation in a bipartite quantum system has been defined
as the difference between the sum of the von Neumann
entropies of the two subsystems and that of the whole system, called the \emph{quantum mutual information} \cite{nielsen}
\begin{eqnarray}
\mathcal{I}_{tot}(\rho)=\mathcal{S}(\rho_A)+\mathcal{S}(\rho_B)-\mathcal{S}(\rho),
\end{eqnarray}
where $\mathcal{S}(\rho)=-\tr (\sigma \ln \sigma) $ and $\rho_A$(and $\rho_B$) is the reduced density operator.
To explore the influence by the correlations outside of the entanglement, we choose a class of two-parameter states as
\begin{eqnarray}\label{rhoIt}
\rho_{I_{tot}}=\rho_{ansatz}|_{y=0,\mathcal{C}= f (\mathcal{N})},
\end{eqnarray}
where $f(x)=-x/2+\sqrt{x+5 x^2/4}$.
 Here, actually, the choice of the function $f(x)$ is quite arbitrary only requires $f(x)\geq x$ when $x\in[0,1]$.
With the above form of the function, the region of the states $\rho_{I_{tot}}$ in (\ref{rhoIt}) in the $\eta$-$\mathcal{N}$ plane is large enough,
and the influence of the quantum mutual information can be shown clearly.
In Fig. \ref{fig3} (c), one can notice for a given $\mathcal{N}$, of course the concurrence $\mathcal{C}$ is fixed, the SoDE decreases with the $I_{tot}$ increasing. That is, the correlations other than the entanglement described by the concurrence also reduce the SoDE under a noise environment.

%
%
%
%
%


\section{Three-qubit pure states}\label{3qubit}

In this section, we explore the SoDE of the three-qubit pure states, whose entanglement properties can be described by the invariants in (\ref{I3qubit}).
For a three-qubit system, there are two inequivalent classes of
genuine tripartite entanglement, which are the GHZ state  and the
W state
\begin{eqnarray}\label{GWstate}
&&|GHZ\rangle=\frac{1}{\sqrt{2}}\bigr(|000\rangle+|111\rangle \bigr), \nonumber\\
&&|W\rangle=\frac{1}{\sqrt{3}}\bigr(|001\rangle+|010\rangle+|100\rangle\bigr).
\end{eqnarray}
They can't be transformed each other by stochastic local operations
and classical communication (SLOCC) \cite{dur1}. The GHZ state
possesses a maximum tripartite entanglement characterized by the
three-tangle \cite{coffman}, for which case $\tau(|GHZ\rangle)=1$,
but the residual bipartite entanglement of GHZ state is zero.
However, the W state possesses zero three-tangle, with
$\tau(|W\rangle)=0$, and maximizes the residual bipartite
entanglement.

In \cite{zhao}, the authors find the most robust symmetrical three-qubit pure states under the partially
depolarizing channel are the GHZ-like states [see below Eq. (\ref{GWlike})], and ascribe the robustness to the three-tangle.
Our question is whether the conclusion is established when we consider the SoDE as the signature for the stability of the entanglement.
Furthermore, what role do the five entanglement invariants play in the SoDE respectively?

\subsection{Symmetrical states}
To answer the first question, we start from the symmetrical pure
states, which are invariable under the permutations of the three
particles.
In this part, the subscripts of the negativity and the SoDE are also omitted.
Generally, an arbitrary symmetrical three-qubit entangled
pure state can be written as \cite{zhao}
\begin{eqnarray}\label{SPureTh}
|\Phi\rangle=t_1|000\rangle+t_2|W\rangle+t_3|W'\rangle+t_4|111\rangle,
\end{eqnarray}
where $t_i$ $(i=1,2,3,4)$ is the complex constant,
$\sum_i{|t_i|^2}=1$,
and $|W'\rangle=(\sigma_x \otimes\sigma_x\otimes\sigma_x)|W\rangle$ with $\sigma_x$ being the first Pauli operator.
The first three entanglement invariants satisfy
$\mathcal{I}_1=\mathcal{I}_2=\mathcal{I}_3=1-\frac{\mathcal{N}^2}{2}$,
where $\mathcal{N}$ is the negativity between one qubit and its complementary two-qubit subsystem.
Only three invariants in the symmetrical states are independent.


\begin{figure}
\subfigure[]{\includegraphics[width=7cm]{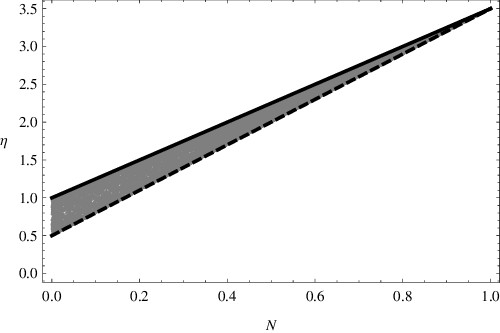}}
\subfigure[]{\includegraphics[width=7cm]{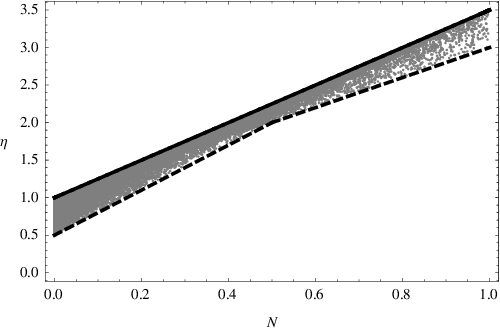}}
 \caption{Plot of (a) 30000 randomly generated
 three-qubit symmetrical pure states; (b) 30000 randomly generated three-qubit
 arbitrary pure states in the $\eta-\mathcal{N}$ plane. The curves show the upper bounds (solid) and the lower bounds (dashed) in the two cases.
 }
\label{fig4}
\end{figure}

As the two-qubit arbitrary states case, we can determined the bounds of the
SoDE in the present case. The region of the symmetrical three-qubit pure states in the $\eta$-$\mathcal{N}$ plane
is shown by the randomly generated states in
Fig. \ref{fig4} (a). Following the approach in the case of the arbitrary two-qubit states,
we obtain the two frontier states on the lower and upper bounds are the GHZ-type states and the W-type states respectively
\begin{eqnarray}\label{GWlike}
&&|G\rangle=\sqrt{a}|000\rangle+\sqrt{1-a}|111\rangle, \nonumber\\
&&|J\rangle=\sqrt{b}|W\rangle+\sqrt{1-b} |111\rangle ,
\end{eqnarray}
where $a,b\in[0,1]$.  They can be proved strictly by using the main result of the present subsection given in Eq. (\ref{Sytotal}).
We remark that the states $|J\rangle$ are not equal to the \emph{W-like states} in \cite{zhao} under LU transformation.

For the state of $|G\rangle$, the partial transpose over the first qubit is $\rho^{T}_{G}=a |000\rangle \langle 000| + \sqrt{a(1-a)}(|011\rangle \langle 100|+|100\rangle \langle 011|) +(1-a)|111\rangle \langle 111|$. The eigenstate with the negative eigenvalue can be obtained as $| \phi_G^{-}
\rangle = (|100\rangle-|011\rangle)/\sqrt{2}$, and the ones with zero eigenvalue are $|001\rangle$, $|010\rangle$, $|101\rangle$ and $|110\rangle$.
The corresponding matrix in (\ref{eta0}) is positive semidefinite, which is written
 in the basis $\{|001\rangle,|010\rangle,|101\rangle,|110\rangle \}$ as
\begin{eqnarray}
\sigma^T_{G,0}=\left[
\begin{array}{cccc}
 \frac{a}{2}& 0 &0    &0\\
 0 &  \frac{a}{2} & 0 &0\\
 0 & 0 &  \frac{1-a}{2} &0\\
0&0&0& \frac{1-a}{2}
\end{array}
\right].
\end{eqnarray}
Therefore, the eigenvectors with the zero
eigenvalue do not have contribution to the SoDE, $ \eta^{(0)}(|G\rangle) = 0$. Similarly,  $ \eta^{(0)}(|J\rangle) = 0$.
Their SoDE are only determined by the responses of the negative eigenvalues under the noise channels,
and can be derived as
\begin{eqnarray}\label{SoDEGJ}
 \eta(|G\rangle)=3\mathcal{N}+\frac{1}{2}, \ \ \ \
\eta(|J\rangle)=\frac{5}{2}\mathcal{N}+1,
\end{eqnarray}
which are the lower and upper bounds respectively.
Hence, in the viewpoint of the SoDE, the GHZ-type states are more robust than the other symmetrical states, which fits with
the result given in \cite{zhao}.

However, it is very attractive to us that whether the robustness comes from the three-tangle.
We can obtain the invariants in the two families of states as
\begin{eqnarray}
\mathcal{N}(|G\rangle)=2 \sqrt{(1-a) a},&& \mathcal{N}(|J\rangle)=\frac{2}{3} \sqrt{2} \sqrt{b (3-2 b)}, \nonumber \\
\tau (|G\rangle)=4 (1 - a) a,&& \tau (|J\rangle)=\frac{16 \sqrt{1-b} b^{3/2}}{3 \sqrt{3}},\\
\mathcal{I}_4 (|G\rangle)=1-3 a+3 a^2,&& \mathcal{I}_4 (|J\rangle)=1-3 b+4 b^2-\frac{16 b^3}{9}. \nonumber
\end{eqnarray}
%
For a given value of negativity, they satisfy $\tau (|G\rangle) \geq \tau (|J\rangle)$ and $\mathcal{I}_4 (|G\rangle) \geq \mathcal{I}_4 (|J\rangle)$, in which the equalities hold when $\mathcal{N}=0$ or $\mathcal{N}=1$.
 This suggests that there exists another candidate, the fourth invariant $\mathcal{I}_4$, for the main factor affecting
the stability of the negativity.

To answer the above question, we consider the hybrid states
of $|G\rangle$ and $|J\rangle$, as
\begin{eqnarray}\label{SPureGJ}
|\Upsilon\rangle=c_1 |000\rangle+ c_2|W\rangle+ c_3|111\rangle,
\end{eqnarray}
where $c_j$ $(j=1,2,3)$ are real and $\sum_j c_j^2=1$.
Its independent entanglement invariants (regardless of the normalization condition) can be derived as
\begin{eqnarray}\label{invts}
\mathcal{N}&=&\frac{2}{3}\sqrt{2c_2^4+9c_1^2c_3^2+6c_2^2c_3^2},\nonumber\\
\mathcal{I}_4&=&c_1^6+3c_1^4c_2^2+\frac{2c_2^6}{9}+c_2^4c_3^2+c_3^6 +\frac{2}{3}
c_1^2c_2^3(3c_2+\sqrt{3}c_3), \nonumber\\
\tau&=&\frac{4}{9}\biggr|c_3(4\sqrt{3}c_2^3+9c_1^2c_3)\biggr|.
\end{eqnarray}
%
%
%
%
%
%
Transposing the states $|\Upsilon\rangle \langle \Upsilon |$ partially on the first qubit,
we derive the unique eigenvector with the negative eigenvalue as
%
\begin{eqnarray}
|\psi^-\rangle&=&
 \frac{3\mathcal{N}+4c_2^2}{6c_1c_3}|000\rangle-\frac{c_2}{\sqrt{3}c_3}|001\rangle-\frac{c_2}{\sqrt{3}c_3}|010\rangle \nonumber\\
  && +
 \frac{3\mathcal{N}+6c_1^2+4c_2^4}{2\sqrt{3}c_1c_2}|011\rangle -\frac{\sqrt{3}\mathcal{N}+2\sqrt{3}c_3^2}{c_2c_3}|100\rangle\nonumber\\
 && -\frac{3\mathcal{N}+2c_2^2+6c_3^2}{6c_3}|101\rangle -\frac{3\mathcal{N}+2c_2^2+6c_3^2}{6c_1c_3}|110\rangle \nonumber\\
  &&+|111\rangle, 
\end{eqnarray}
which is non-normalized.
 Under the
depolarizing channel in an infinitesimal time, the form
of $\sigma^{T}$ can be derived directly.
 Substituting it and the eigenvector into (\ref{pert}), we obtain the
SoDE in terms of the coefficients $c_j$ as $\eta(c_1,c_2,c_3)$.
In this case, $\eta^{(0)}=0$ means the eigenvectors with the zero eigenvalue do not have contribution to the SoDE.
Considering the relations $\tau=\mathcal{N}^2$ and $\mathcal{I}_4=1-3\mathcal{N}^2 /4$, satisfied by
the GHZ-type states, we suppose the SoDE of the states $|\Upsilon\rangle$ can be written as
\begin{eqnarray}
\eta=3\mathcal{N}+\frac{1}{2} + \bigr(\tau-\mathcal{N}^2\bigr) X+ \bigr(\mathcal{I}_4-1+\frac{3}{4}\mathcal{N}^2 \bigr)Y,
\end{eqnarray}
where $X$ and $Y$ are two undetermined functions of $\mathcal{N}$, $\tau$ and $\mathcal{I}_4$.
Substituting the invariants (\ref{invts}) into the above expression, and comparing it with $\eta(c_1,c_2,c_3)$, we obtain the
form of SoDE for the state $|\Upsilon\rangle$ as
\begin{eqnarray}\label{Sytotal}
\eta=\frac{32-32\mathcal{I}_4-12\mathcal{N}^2+84\mathcal{N}^3+69\mathcal{N}^4+3\tau^2}{24\mathcal{N}^2(\mathcal{N}+1)}.
\end{eqnarray}
Very fortunately, we find the relation is also satisfied by the general symmetry states (\ref{SPureTh}).


From the above relation between the SoDE and the entanglement invariants,
one can clearly notices that, $\eta$ decreases with the increase of $\mathcal{I}_4$,
but increases as the three-tangle increasing.
In other words, the entanglement among the three qubits, quantified by the three-tangle, can enhances
the SoDE, but the two-qubit correlations described by $\mathcal{I}_4$ reduces the speed of negativity.
Therefore, it is $\mathcal{I}_4$ but not the three-tangle $\tau$ making the GHZ-type states to be the most robust symmetric pure states.
In the analysis of the Ref. \cite{zhao}, the states they considered have no more than two independent parameters,
 which muddles up  the influences of  $\mathcal{I}_4$ and $\tau$.

%

\subsection{General states}

The discovery of the analytic expression of the relation between the SoDE and
the entanglement invariants in symmetrical states in (\ref{Sytotal}) reveals the SoDE is an effective tool
to explore the role of different entanglement in the stability of multipartite entanglement.
This motivates us to extend it to more general case, and study the influences of the other invariants on the SoDE.
%

The general form of three-qubit pure states is given by
\begin{eqnarray}\label{GPure}
|\Psi\rangle &=& c_0|000\rangle+c_1|001\rangle+c_2|010\rangle+c_3|011\rangle \nonumber\\
&&+c_4|100\rangle+c_5|101\rangle+c_6|110\rangle+c_7|111\rangle,
\end{eqnarray}
where $c_i(i=0,1,2,...,7)$ are complex constants and $\sum_i{|c_i|^2}=1$.
Without loss of generality, the analysis can be restricted to the negativity between
the first qubit and the subsystem containing the other two qubits.

Following our schedule in the symmetrical states case, we first derive the bounds of
the SoDE for the arbitrary pure states. Plotting the randomly generated three-qubit pure
 states in the $\eta_1$-$\mathcal{N}_1$ plane, we find only the lower bound are different with
the one of the symmetrical states in the region of $\mathcal{N}_1 > 1/2$.
And the altered lower bound corresponds to nothing but the result of the pure two-qubit state
in (\ref{SoDE2bit}). Thus, the frontier states on the lower bound of the SoDE for three-qubit
 pure states are the GHZ-type states when $\mathcal{N}_1 \leq 1/2$, and the direct product states
of an entangled state containing the first qubit and a single partite state of the remaining qubit,
such as $|\mu\rangle=|\psi(\theta)\rangle|0\rangle$ with the $|\psi(\theta)\rangle$ defined in (\ref{PureTh}).
Besides the symmetric pure states $|J\rangle$, we also find the frontier states on
the upper bound contain the family of states as
%
\begin{eqnarray}
|\Pi\rangle=\alpha|0\rangle|\psi(\theta_0)\rangle+\beta|1\rangle|\psi'(\theta_1)\rangle,
\end{eqnarray}
where $|\alpha|^2+|\beta|^2=1$, $|\psi'(\theta)\rangle=(I\otimes\sigma_x)|\psi(\theta)\rangle$
and $(\theta_0-\pi/4)(\theta_1-\pi/4)=0$.




To realize the ultimate aim of the SoDE for the arbitrary pure states,
we consider the subclass of states of $\Psi$ as
\begin{eqnarray}\label{Tlower}
|\Lambda\rangle = c_0|000\rangle+c_1|001\rangle+c_6|110\rangle,
\end{eqnarray}
where $c_i$ are restricted in the real numbers. It can be regarded a hybrid state of the
two frontier states on the lower bound of the SoDE, which becomes the GHZ-type state
when $c_0=0$ and returns to $|\mu\rangle$ with $c_1=0$.
%
%
%
 Its five entanglement
invariants can be derived as
\begin{eqnarray}\label{invtsLam}
\mathcal{I}_1&=&\mathcal{I}_2=1-\frac{\mathcal{N}_1^2}{2},\nonumber\\
\mathcal{I}_3&=&1-2c_1^2 c_6^2 ,\\
\mathcal{I}_4&=&c_0^6+3 c_0^4 c_1^2+3 c_0^2 c_1^4+c_1^6+c_6^6,\nonumber\\
\tau&=&4c_1^2c_6^2,\nonumber
\end{eqnarray}
where $\mathcal{N}_1=2\sqrt{(c_0^2+c_1^2)c_6^2}$.
In the the perturbation approach,
%
the first term of the SoDE for $|\Lambda\rangle$ the can be gotten as
\begin{eqnarray}\label{Tlower1}
\eta_1^{(-)}\bigr(|\Lambda\rangle\bigr)=\frac{5}{2}\mathcal{N}_1+1-\frac{(1-\mathcal{N}_1)(1-\mathcal{I}_3)}{\mathcal{N}_1^2}.
\end{eqnarray}
There exist four eigenstates of the partial transposed density matrix of $|\Lambda\rangle$ with the zero eigenvalue,
 which span the degenerate subspace $\{ |111\rangle, c_1 | 100\rangle-c_0 |101 \rangle, |011\rangle, c_1 |000\rangle- c_0 |001\rangle \}$.
And the matrix $\sigma^{T_1}_0$ defined in (\ref{eta0}) in the subspace is given by
%
%
\begin{eqnarray}
\sigma^{T_1}_0=\left[
\begin{array}{cccc}
 \frac{c_6^2}{2} & 0 & 0 & 0 \\
 0 & \frac{c_1^2 c_6^2}{2 \left(c_0^2+c_1^2\right)} & \frac{c_0^2 c_6}{2 \sqrt{c_0^2+c_1^2}} & 0 \\
 0 & \frac{c_0^2 c_6}{2 \sqrt{c_0^2+c_1^2}} & \frac{c_1^2}{2} & 0 \\
 0 & 0 & 0 & \frac{c_0^2+c_1^2}{2}
\end{array}
\right].
\end{eqnarray}
The criterion for the matrix $\sigma^{T_1}_0$ having a negative eigenvalue is $c_1^4c_6^2-c_0^4c_6^2<0$.
From the relations in (\ref{invtsLam}), we find it can be replaced by $\Theta=(\mathcal{I}_2-\mathcal{I}_3)^2-\tau^2/4>0$,
which is verified as the universal criterion for nonzero $\eta_1^{(0)}$ for arbitrary three-qubit pure  states.
By using the relations in (\ref{I3qubit}),
the criterion can be rewritten as $|\tau_{12}-\tau_{13}|-\tau > 0$,
where $\tau_{ij}=\mathcal{C}_{ij}^2$ are the two-tangle.
The second part of the SoDE for $|\Lambda\rangle$ can be written as
\begin{eqnarray}\label{Tlower2}
\eta_1^{(0)}\bigr(|\Lambda\rangle\bigr)=\left\{
\begin{array}{lr}
0,     \; &  \Theta \leq 0 \; , \\
\frac{\sqrt{(1-\mathcal{I}_3)^2+\mathcal{N}_1^2 \Theta }-(1-\mathcal{I}_3)}{\mathcal{N}_1^2},
\; & \Theta > 0 \;.
\end{array}
\right.
\end{eqnarray}


On the other hand, we calculate a class of states without the three-tangle as
%
%
\begin{eqnarray}
|\Omega\rangle=c_0|000\rangle+c_1|001\rangle+c_2|010\rangle+c_4|100\rangle,
\end{eqnarray}
where$\sum_j{|c_j|^2}=1$ with $j=0,1,2,4$. It can be considered as a generalization
of the W-like states in \cite{zhao}
%
Similarly, for $|\Omega\rangle$, following the perturbation approach and replacing $c_j$ by the invariants,
we get the SoDE
\begin{widetext}
\begin{eqnarray}\label{Omeq}
\eta_1 \bigr( |\Omega\rangle \bigr)&=&\frac{-16-32\mathcal{I}_4+36\mathcal{N}_1^2+84\mathcal{N}_1^3+57\mathcal{N}_1^4-12(\mathcal{I}_2-\mathcal{I}_3)^2
+12(\mathcal{I}_2+\mathcal{I}_3)(2-\mathcal{N}_1^2)}{24\mathcal{N}_1^2(\mathcal{N}_1+1)} \nonumber \\
&&-\frac{\sqrt{M^2+(\mathcal{I}_2-\mathcal{I}_3)^{2}\mathcal{N}_1^2}-M}{\mathcal{N}_1^2}. \ \
\end{eqnarray}
where
$M=\frac{1}{3}(5-3\mathcal{I}_1-3\mathcal{I}_2-3\mathcal{I}_3+4\mathcal{I}_4)=\frac{1}{6}(4-6\mathcal{I}_2-6\mathcal{I}_3+8\mathcal{I}_4+3\mathcal{N}_1^2)
$.

%
%
%

Based on an overall consideration of the analytic results of SoDE in the states $|\Phi\rangle$, $|\Lambda\rangle$ and $|\Omega\rangle$, we guess
the form of SoDE for general three-qubit pure states to be
\begin{eqnarray}\label{eta3qubit}
\eta_1\bigr( | \Psi \rangle \bigr)&=&\frac{-16-12\Theta-32\mathcal{I}_4+36\mathcal{N}_1^2+84\mathcal{N}_1^3+57\mathcal{N}_1^4
+12(\mathcal{I}_2+\mathcal{I}_3)(2-\mathcal{N}_1^2)}{24\mathcal{N}_1^2(\mathcal{N}_1+1)} \nonumber \\
&&-\left\{
\begin{array}{lr}
0,     \; &  \Theta\leq0 \;;\\
\frac{\sqrt{M^2+\mathcal{N}_1^2\Theta}-M}{\mathcal{N}_1^2}, \; &
\Theta>0 \;.
\end{array}
\right.
\end{eqnarray}
\end{widetext}
By the numerical validation, the above unified form almost is fulfilled
by most all the three-qubit pure states.
Namely, we take
$\Delta\eta=|\eta_1-\eta_m|$, where $\eta_m$ is the numerical solution by the finite difference method with $\Delta t=10^{-9}$.
We generate $200000$ sets of data for the $\Delta\eta$, which show
 that $\Delta\eta$ is less than $10^{-5}$ when
$\mathcal{N}_1\geq10^{-5}$. When $\mathcal{N}_1<10^{-5}$,
since the emergence of the phenomenon of ESD, the finite difference
method is no longer suitable.

 From the form in (\ref{eta3qubit}), one can notice the conclusions about the influences of $\mathcal{I}_4$
 and $\tau$ on the SoDE drawn from the symmetrical states also exist in the general case.
 The sum of the invariants $\mathcal{I}_2$ and  $\mathcal{I}_3$ can enhance the SoDE.
 In the other words, the decay of the negativity between the first qubit and the subsystem containing the
 other two qubits, can be restrained by the other two negativity with the coequal sense.
 The unbalance between the second and the third qubits, shown by the differences $(\mathcal{I}_2-\mathcal{I}_3)^2$ and $|\tau_{12}-\tau_{13}|$,
 can reduce the  SoDE of the first qubit.
Especially when $|\tau_{12}-\tau_{13}|>\tau$, the vanishing negativity can be partially brought back,
 which reflected in nonzero $\eta_1^{(0)}$.


\section{Multiqubit states}\label{mqubit}

For the multiqubit system, in which the SoDE will be more complex,
due to the absence of a uniform measure of entanglement, we only
study some special symmetrical states.

The $k$-qubit GHZ-type states have been investigated widely as a family of
exemplary states in the topic about the decay of entanglement in multiparty systems.
It is given by
\begin{eqnarray}\label{GHZtype}
|G\rangle_k=\alpha|0\rangle^{\otimes k}+\beta|1\rangle^{\otimes k},
\end{eqnarray}
where $\alpha,\beta$ are complex constant and $|\alpha|^2+|\beta|^2=1$.
%
For the case of $k\geq3$, the SoDE under the local depolarizing channel can be calculated directly as
\begin{eqnarray}\label{SoDEkG}
\eta(|G\rangle_k)=k \mathcal{N}+\frac{1}{2},
\end{eqnarray}
where the negativity $\mathcal{N}=2|\alpha\beta|$.
And, the alternative robustness is given by
\begin{eqnarray}\label{robGk}
\mathcal{R}_{\eta}(|G\rangle_k)=1-\Exp \biggr(-\frac{2 \mathcal{N}}{2k \mathcal{N}+1}\biggr).
\end{eqnarray}
It is interesting to notice the SoDE given by (\ref{SoDE2bit}) of the two-qubit GHZ-type states
isn't contained in the form of (\ref{SoDEkG}).
Based on the results in (\ref{SoDEGJ}), we present a possible explanation as, the slop in the SoDE (\ref{SoDE2bit}) of the two-qubit state comes from
its similarity with the GHZ-type states, but the intercept is related with its property of the W-type states,
 since they are the two different generalizations of the two-qubit pure states to the three-qubit system.
In the limit of large number of particles $k\rightarrow + \infty$, the robustness (\ref{robGk}) and the corresponding characteristic time
 approach $ \mathcal{R}_{\eta} \rightarrow 1/k$ and $ T^* \rightarrow 1/k$,
which conforms to the time at which such entanglement becomes arbitrarily small \cite{aolita}.

%
%
%
%
%

\begin{figure}
\subfigure[]{\includegraphics[width=7cm]{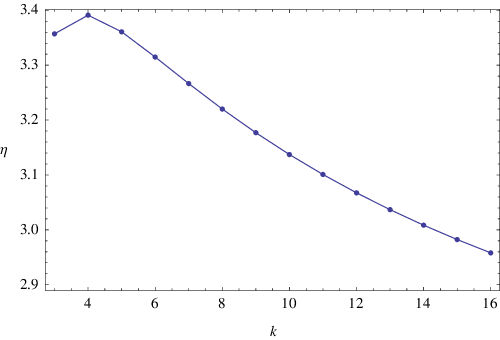}}
\subfigure[]{\includegraphics[width=7.5cm]{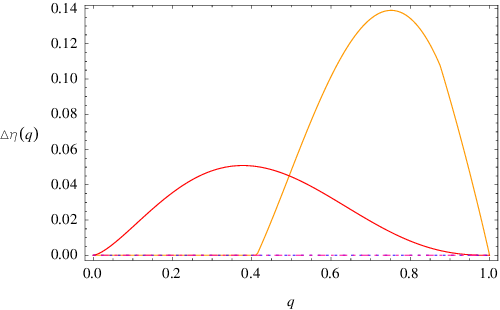}}
 \caption{(a) Plot of the relation between the SoDE and the number of qubit for $|W\rangle_k$.
 (b) Plot of the maximal influence of $\varphi$ on the SoDE of the states
$|Z(q,\varphi)\rangle_k$ for $k=3$ (solid), $k=4$ (dashed) and  $k=5$ (dot-dashed).
 }
\label{fig5}
\end{figure}

The second class of $k$-qubit entangled states we considered are the W states
\begin{eqnarray}\label{wstate}
|W\rangle_k=\frac{1}{\sqrt{k}}\bigr(|00...01\rangle+|00...10\rangle+...+|10...00\rangle\bigr),
\end{eqnarray}
whose negativity is $\mathcal{N}=2\sqrt{k-1}/k$. %
The corresponding SoDE can be obtained as
\begin{eqnarray}\label{SoDEwstate}
\eta(|W\rangle_k)=\frac{(k+2)\sqrt{k-1}+2(k-1)-(k-2)\sqrt{k-2}}{k}. \ \
\end{eqnarray}
It is plotted in Fig. \ref{fig5} (a), one can find that the SoDE
$\eta$  increases with $k$ for $2\leq k\leq4$,
but decreases with $k$ when $k\geq5$. However, the robustness
of $|W\rangle_k$ only decrease with $k$.
When the number of particles $k\rightarrow + \infty$, the negativity  $\mathcal{N}\rightarrow 2/\sqrt{k}$,
and the SoDE approaches a constant $\eta(|W\rangle_k) \rightarrow 2$.
 And then the alternative robustness of $|W\rangle_k$ is given by $ \mathcal{R}_{\eta}(|W\rangle_k) = 1-\Exp (-  \mathcal{N}/\eta  ) \rightarrow 1/\sqrt{k}$.
It is worth mentioning the result in \cite{chaves2010robustness} that, for the $k$-qubit system initial with the W states under the amplitude-damping channel, the negativity of the least balanced partitions  decays with $1/\sqrt{k}$.

%
%



When the number of qubits  $k=3$, the GHZ states $|GHZ\rangle_k= (|0\rangle^{\otimes k}+ |1\rangle^{\otimes k})/\sqrt{2}$
 and the W states in (\ref{wstate}) become the two frontier states (\ref{GWlike}) for the symmetric case with $a=1/2$ and $b=1$.
  And the negativity of their superposition
\begin{eqnarray}
|Z(q,\varphi)\rangle_k=\sqrt{q}|GHZ\rangle_k-e^{i\varphi}\sqrt{1-q}|W\rangle_k,
\end{eqnarray}
does not depend on the phase factor $\varphi$ \cite{zhao}.
 Therefore, the influence of $\varphi$ on the SoDE is related with the entanglement properties described by $\tau$ and $\mathcal{I}_4$.
Since, for arbitrary values of $k$, the negativity of $|Z(q,\varphi)\rangle_k$
\begin{eqnarray}
\mathcal{N}=\frac{\sqrt{(-k^2+6k-4)q^2+2(k^2-5k+4)q+4(k-1)}}{k}. \  \
\end{eqnarray}
 also only depends on the parameter $q$,
we can consider the maximal effect of $\varphi$
\begin{eqnarray}
\Delta\eta(q)=\max\{\eta[|Z(q,\varphi_a)\rangle_k]-\eta[|Z(q,\varphi_b)\rangle_k]\},
\end{eqnarray}
as a characteristic quantity to indicate the affection of the entanglement besides the negativity.
In Fig. \ref{fig5} (b), we plot the relation between $\Delta\eta(q)$ and $q$ for $k=3,4,5$.
It is shown that, when $k=3$, the phase  $\varphi$  presents a significant affection on the SoDE for
arbitrary $q$. But, for the case of $k=4$, the influence is not distinct when $q<0.4$, where the proportion of
the GHZ state in $|Z(q,\varphi)\rangle_k$  is less than the one of the W state.
When $k=5$, the values of $\Delta\eta(q)$ can hardly be seen in the Fig. \ref{fig5} (b).
With a numerical simulation, we find the amount of $\Delta\eta(q)$ is no more than $10^{-4}$, for $k=5$ and $6$.
That is, the SoDE of the states $|Z(q,\varphi)\rangle_k$ are almost entirely on the negativity when $k$ becomes large.

By contrast, one can choose the GHZ-type states (\ref{GHZtype}) with the negativity $\mathcal{N}=2\sqrt{k-1}/k$, and derive
the difference between the values of SoDE in (\ref{SoDEkG}) and (\ref{SoDEwstate}).
It is easy to find the difference increases with increasing number of qubits, when $k\geq4$.
This reveals, the affection of the entanglement besides the negativity on the SoDE becomes more prominent as the number $k$ increases.
According to these results, the family of the states $|Z(q,\varphi)\rangle_k$ is not a good sample to explore the role of different entanglement
components in the stability of the entanglement in the systems with a large number of particles.

\section{Conclusion and Discussion}\label{conclu}
In the present study, we investigate the dynamical properties of
entanglement of multiqubit systems under local partially
depolarizing channels. In this model, each qubit is independent
couples with its own environment. Our main concern is the bipartite
entanglement between one qubit and its complementary subsystem,
measured by negativity. The relations between the SoDE and the
entanglement properties in an entangled state are explored.

In two-qubit system, for the pure states, we get the analytical
expression of the SoDE, which is determined completely by the
negativity. For the arbitrary states, using the form of the ansatz
states, we gain the upper and lower bounds of the SoDE. The pure
states has the maximal SoDE, and the mixture of mutually orthogonal
a Bell state and a separable pure state as (\ref{rhom}) achieves the
minimum of SoDE. The minimum and the maximum of concurrence are
shown to be the necessary conditions for the two bounds respectively.
With the aid of some classes of two-parameter states, we find the
SoDE can be reduced by both the concurrence and the total
correlation.

 In three-qubit system, we derive the analytical expression of the SoDE
in terms of the entanglement invariants for arbitrary pure states.
The GHZ-type states are shown to be the most robust, which is consistent with
the result in \cite{zhao}. However, by the relation between the SoDE and
 the invariants, we find the main reason for the robustness in the GHZ-type states
 is the two-qubit correlation described by the fourth invariant $\mathcal{I}_4$.
On the contrary, the three-tangle among the whole system can enhance
 the speed of negativity under the local depolarizing channel.
The decay of the negativity can be restrained by the other two negativity with the coequal sense.
The unbalance between two qubits can reduce the SoDE of the remaining one  in the three-qubit system.
Especially when $|\tau_{12}-\tau_{13}|>\tau$, the vanishing negativity of the first qubit can be partially brought back.

At last, we study $k$-qubit system with the aid of some families of exemplary states.
For the $k$-qubit systems in an arbitrary superposition
of GHZ state and W state, the influence of the entanglement outside the negativity becomes less noticeable when $k$ increases to five or six.
In the limit of large number of particles, for the GHZ-type states, the characteristic time and the corresponding
robustness defined in (\ref{aRob}) base on the SoDE, is inversely proportional to $k$.
This coincide with the result of the time at which such entanglement becomes arbitrarily small given in \cite{aolita}.
Under the same condition, the robustness of the $k$-qubit W states approaches $1/\sqrt{k}$.
 A similar behavior has been reported in \cite{chaves2010robustness}, exhibited by the the negativity of the least balanced partitions, in the
 $k$-qubit system initial with the W states under the amplitude-damping channel.
These results show the effectiveness of the SoDE and the corresponding alternative robustness to
quantize the stability of the entanglement in multipartite quantum systems.



Finally, we briefly discuss the universality of our definitions and results, especially about the choices of the entanglement measures and the noise channels.
The main reason for adopting the negativity as the entanglement measure in our present work about the SoDE is its computability, and fatherly its speed being also computable effectively with the aid of perturbation theory.  A computable quantity always has more advantages for physicists.
Because of the necessity and sufficiency of $\mathcal{N}>0$ for entanglement in pure states, the SoDE defined in (\ref{eta}) and the related alternative robustness can be used to explore the stability of the entanglement in arbitrary dimensional multiparty pure states.
 In addition, although $\mathcal{N}>0$ only is the sufficiency condition for entanglement in mixed states with the dimension larger than six,
the speed of negativity can still be considered as a signature for stability of the entanglement in the states with a nonzero $\mathcal{N}$.
Meanwhile, the robustness related with the ESD noise parameter is suspect, because its practical computation is usually based on the PPT criterion.

In principle, on can choose any measures of entanglement and study the corresponding SoDE.
We conjecture that, in the multiparty system with pure states, the qualitative conclusions for other entanglement measure  are similar with the ones for negativity.

 Taking the concurrence in (\ref{ConM}) for instance, one can define its speed as the form in (\ref{eta})
\begin{eqnarray}\label{etaC}
\eta^{C}=\biggr| \frac{d \mathcal{C}}{d t} \biggr|_{t=0}.
\end{eqnarray}
From the relation in (\ref{NboundC}), it is directly to find the speed of the concurrence for a pure state in multiqubit system satisfies
\begin{eqnarray}\label{etaCN}
\eta^{C} \leq \eta^{N},
\end{eqnarray}
where $\eta^{N}$ denotes the corresponding SoDE in (\ref{eta}).
Especially, for the two-qubit case,
\begin{eqnarray}\label{etaC2bit}
\eta^{C} = 2\mathcal{C}   +1 = \eta^{N}.
\end{eqnarray}
And, for the $k$-qubit GHZ-type states
\begin{eqnarray}\label{etaCGHZ}
 k\mathcal{C}+\frac{\mathcal{C}}{2} \leq \eta^{C} \leq k\mathcal{C}+\frac{1}{2}  = \eta^{N},
\end{eqnarray}
where the upper bound comes from (\ref{etaCN}), and the lower bound can be derived as following.
After passage through the depolarizing channel, the final state can be divided as
\begin{eqnarray}\label{GHZs}
\rho_s^{G}=[1-\delta(s)]\rho_{e}(s)+\delta(s) \rho_{d}  (s),
\end{eqnarray}
where $\rho_{e}(s)$ is a state in the space of $\{|0\rangle,|1\rangle \} \otimes \{ |0\rangle^{\otimes(k-1)},|1\rangle^{\otimes(k-1)} \}$,
$\rho_{d} (s)$ is a separable diagonal state, and they together with  the non-negative coefficient $\delta(s)$ depend on the noise parameter $s$.
By the convexity of concurrence,
\begin{eqnarray}\label{CGHZs}
\mathcal{C}(\rho_s^{G})\leq[1-\delta(s)]\mathcal{C}[\rho_{e}(s)]
\end{eqnarray}
which leads to the lower bound in (\ref{etaCGHZ}). Moreover, based
on the bounds in (\ref{etaCGHZ}), we can obtain the corresponding
alternative robustness
$\mathcal{R}^{C}_{\eta}=1-\Exp(-\mathcal{C}/\eta^C) \rightarrow
1/k$, when $k \rightarrow \infty$, which is the same as the result
for negativity. Then, the above results left a interesting question
for us: Can we find a explicitly expression for $\eta^C$ of the
multiqubit pure states?

However, for the mixed states, the speed of concurrence has a
problem of singularity. For example, in the two-qubit system,
  the form of ansatz states (\ref{ansatz}) is held under the local depolarizing channel. The speed
of concurrence can be obtained as $\eta^C=\sqrt{\frac{b}{a}} \frac{\partial a}{\partial t }+\sqrt{\frac{a}{b}} \frac{\partial b}{\partial t } -  \frac{\partial \gamma}{\partial t } $, which approaches
infinity when $a > 0$ and $b \rightarrow 0$ (or  $a \rightarrow 0$ and $b > 0$).
 This property also exists under other local operations \cite{verstraete2001local}, which holds the form of the ansatz states and drastically enforces full rank of the states when $a=0$ or $b=0$. However the corresponding alternative
robustness $\mathcal{R}^{C}_{\eta}$ has a finite value.

In order to compare with the depolarizing channel, we also calculate
the SoDE of the GHZ-type states $|G\rangle_k$ under the local
dephasing channel, which is described by the Kraus operators as
$E_0=\sqrt{(1+ e^{-t})/2} I$ and $E_1=\sqrt{(1- e^{-t})/2}
\sigma_3$. The SoDE for both the concurrence and the negativity are
\begin{eqnarray}\label{etaPhase}
\eta = k \mathcal{C}=k \mathcal{N},
\end{eqnarray}
which also lead the alternative robustness $\mathcal{R}_{\eta}
\rightarrow 1/k$, with $k \rightarrow \infty$. This is in accord
with the conclusion about the limit of large number of particles in
\cite{PhysRevA.71.032350}.




\begin{acknowledgments}
 We are grateful to the reviewer for valuable comments.
 F.L.Z. thanks Jing-Ling Chen, Fu-Guo Deng and Bao-Kui Zhao for their valuable discussions and encouragement.
 This work is supported by NSF of China (Grant No. 11105097).
\end{acknowledgments}

\bibliography{SpeedDisEntangle}

\end{document}